\documentclass[a4paper,twocolumn]{esapub2005} % European paper
\usepackage{times}
\usepackage{graphicx}
\usepackage{natbib}

\title{ToO observations of GRO J1655-40 in outburst with INTEGRAL}
\author[1]{M. D. Caballero-Garc\'{\i}a}
\affil[1]{LAEFF-INTA   P.O. Box 50727, 28080 Madrid (Spain), E-mail: mcaballe@laeff.inta.es}
\author[2]{E. Kuulkers}
\author[2]{P. Kretschmar}
\affil[2]{ESA/ESAC, Urb. Villafranca del Castillo, PO Box 50727, 28080 Madrid (Spain)}
\author[1]{A. Domingo}
\author[3]{J. M. Miller}
\affil[3]{University of Michigan, Department of Astronomy, 500 Church Street, Dennison 814, Ann Arbor, MI 48105 (USA)}
\author[4]{J. M. Mas-Hesse}
\affil[4]{CAB (CSIC-INTA) P.O. Box 50727, 28080 Madrid (Spain)}

\begin{document}


\maketitle

\begin{abstract}
In this paper we present the results of the analysis of the INTEGRAL data for the black hole transient GRO J1655-40.
The observations consist of four ToO (100 ks each) observations's AO-3 spread from February to April of 2005. We
present here the preliminary spectral analysis of this source between 5 and 200 keV. Also, some comments of the 
light curves obtained during this period are shown.
\end{abstract}

\section{Introduction}

The X-ray transient GRO J1655-40 (also called X-ray Nova Scorpii 1994) was discovered with the Burst and Transient Source 
Experiment (BATSE) on board the 
Compton Gamma-Ray Observatory (CGRO) on 1994 July 27 \citep{zhang94}. The optical counterpart was discovered 
soon after by \citet{bailyn95a}. Subsequent optical studies showed that the system is an LMXB composed of a 
blue subgiant (spectral type F4 IV) as the secondary and a black hole as the primary ($m_{BH}=7.02{\pm}0.22$$M_{\odot}$)
\citep{orosz97a}, located at a distance of 3.2 kpc \citep{tingay95} (see also \citet{foellmi06}). \citet{bailyn95b} (see also \citet{orosz97b} and \citet{hooft98}) 
%ERIK PROPONIA COMO VALOR DISTANCIA A OTRA REF.: FOELLMI ET AL. (2006). ESTUDIARLA.
established the orbital inclination of the system to be ${\simeq}70^{\circ}$. 

%PONER EN EL PAPER:
%GRO J1655-40 is one of a few Galactic X-ray sources known to produce superluminal radio jets (\citet{tingay95} and 
%\citet{hjellming95}). This kind of sources, characterized by the presence of double-lobed radio structures 
%(see also \citet{hannikainen00}) are known collectively as ``microquasars'', since they have 
%properties analogous to the radio-loud active galactic nuclei. 

Galactic black hole binaries (hereafter BHB) show interesting luminosity/spectral properties which appears
to be 
strongly related with the accretion behaviour onto the black hole (hereafter BH). 
Matter accretes onto the BH through an accretion disk, which has nearly Keplerian orbits; the last stable orbit is called ``Innermost Stable Circular Orbit'' 
(i.e. ISCO) \citep{shapiro83} which is located at $R_S=6R_{g}$ (for a Schwarzschild BH; where $R_{g}=GM/c^{2}$ is the 
gravitational radius). Note that not always the matter arrives at this radius; this
depends on the particular state. 

There are five states known in BHB: quiescent, low-hard, intermediate, high-soft 
and very high states in the unification scheme (e.g. \citet{esin97}); the inner radius 
%ERIK PROPONE OTRAS REPRESENTACIONES DEL ESQUEMA  DE LA UNIFICACION: McClintock & Remillard, 2006? estudiarlo!!!
of the accretion disk is thought to decrease from quiescence to VHS. At lower mass accretion rate, corresponding to several percent of the Eddington luminosity,
a BHB usually enters the low-hard (LH) state and at very low accretion rates it reaches the quiescent state, which may 
be just an extreme of the LH state. In both of these states, the spectrum of a BHB is dominated by a hard, nonthermal
power-law component (photon index ${\sim}1.7$); it is most plausibly explained as due to Comptonization of soft photons (from 
a hot optically thick disk) by a hot optically thin plasma. \citet{narayan96} postulated that in these states the disk does 
not extend down to the ISCO, but is truncated at some larger radius and the interior volume is filled by a hot 
($T_{e}{\sim}100$ keV), radiative inefficient, advection-dominated accretion flow or ADAF. 

The Multicolor Disk Model (MCD) (\citet{mitsuda84}, which is an application of the \citet{shakura73} disk) is
used to describe the thermal component that is dominant in the Very High (VH) and High-Soft (HS) states. In 
Figure 4.1 \citep{mcclintock04} the unification model from \citet{esin97} is pictured, using both MCD and ADAF models. It represents
how the geometry of the accretion flow changes as the mass accretion rate from the donor star varies. When the accretion is
low, the flow consists of two zones (disk and ADAF) and for the two states in which the accretion is high, the disk 
extends down to the ISCO. In all five states the disk is bathed in a corona which resembles a continuation to the ADAF. 
Although this model had success in the explanation of evolution of the states, it has important limitations not 
discussed in this paper. 

In this paper, we briefly report on observations of the BHB GRO J1655-40 made by INTEGRAL.
After describing the observations and our
analysis techniques, we present the preliminary results of the spectral fits. We then briefly discuss the results and
end with some concluding remarks.

\begin{figure}
\centering
\vspace{4cm}
\includegraphics[width=1.0\linewidth]{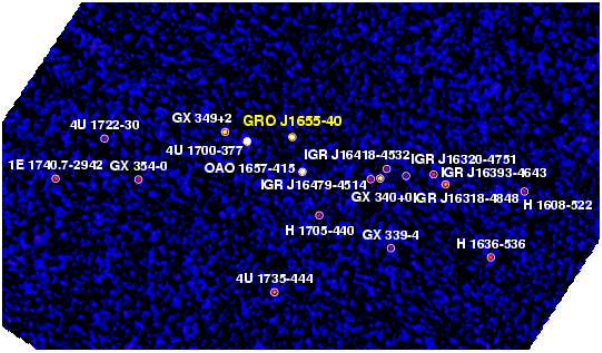}
\caption{Mosaic image (obtained in revolution 295) of the GRO J1655-40 region as seen by ISGRI in the 20-40 keV energy range. Besides the target source, several other high energy sources are visible.\label{mosaic}}
\end{figure}

\section{Observations}

The data were obtained with INTEGRAL and cover the outburst using the instruments SPI, IBIS/ISGRI, JEM-X and OMC. These 
first part of the observations were made during 4 ToO (of 100 ks each) spread from 27 February to 11 April of 2005.
The dithering pattern used during the observations was $5{\times}5$. Data analysis (in the
case of JEM-X, IBIS/ISGRI and SPI) was performed using the standard OSA 5.1 analysis software package available from the 
INTEGRAL Science Data Centre (hereafter ISDC). In the case of SPI, because of the lower angular resolution and crowdedness
of the Field of View (FOV) in ${\gamma}$-rays at this position of the sky ($l$,$b$)=($344.98^{\circ}$,$+2.46^{\circ}$), 
(see Figure \ref{mosaic}),
we used a non-standard procedure in the analysis of the data, described in \citet{deluit05} and \citet{roques05}. For the 
same reason, in the case of OMC, we 
used a non-standard pipeline for extraction of fluxes 
(A. Domingo
, private communication), which will be delivered under OSA 6.0 in the very near future. The data we analized 
come from the observations with P.I.: Miller. In the case of OMC, all the public data from
ISDC were downloaded (this implies only a slight increase of data). In order to avoid large off-axis
angles in the case of JEM-X and because of its reduced FOV ($5^{\circ}$ of diameter), we limited the radius of directions of pointings 
with respect to the GRO J1655-40 position to be $4^{\circ}$ only. In the case of SPI and ISGRI, with large Fully Coded Fields 
of View (FCFOV) ($16^{\circ}{\times}16^{\circ}$ for SPI and $9^{\circ}{\times}9^{\circ}$ for IBIS) this selection of pointings was not applied. 
Summarizing, 199 individual pointings were used for SPI and IBIS/ISGRI, 96 pointings for JEM-X and 66 pointings for OMC. We
ommit the SPI results in the current paper, since the analysis is still in progress.

\section{Light Curves} 

\begin{figure}
\centering
\vspace{4cm}
\includegraphics[width=1.0\linewidth]{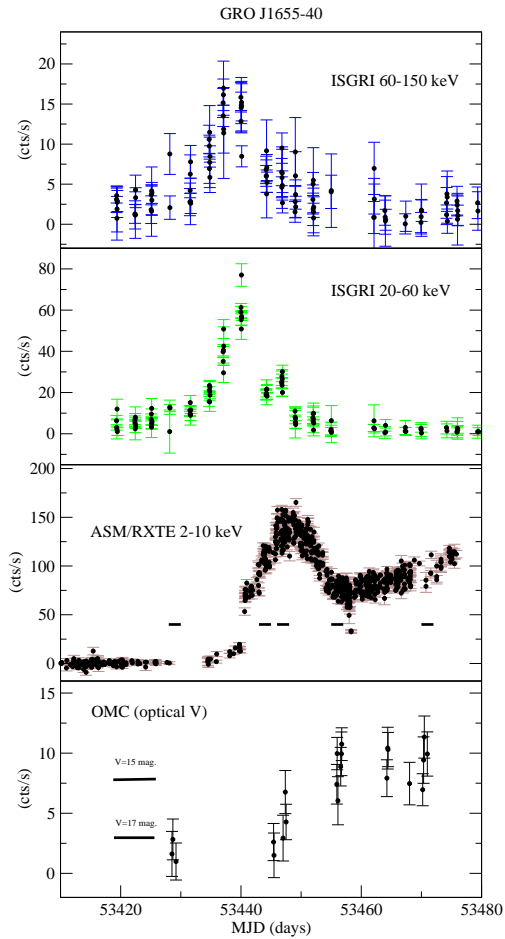}
\caption{Light curves obtained with IBIS/ISGRI from the INTEGRAL Galactic Monitoring Program in two energy
bands (60-150 keV and 20-60 keV), together with OMC (optical). The horizontal lines in the OMC panel show the equivalence
in magnitudes of the fluxes. In the third panel, ASM/RXTE (2-10 keV) light curve is shown in the same period of time. The horizontal
lines indicate the time intervals (one revolution each) over which INTEGRAL spectra were obtained. \label{curves}}
\end{figure}

Figure \ref{curves} shows the GRO J1655-40 light curves obtained by IBIS/ISGRI (from the INTEGRAL Galactic Bulge Monitoring Program) in two energy bands (60-150 keV and 20-60 keV) together with OMC (optical). 
The ASM/RXTE (2-10 keV) light curve in the same period of time
is shown in the same figure. A progressive 
delay in the outburst can be seen in the softer energy range with respect to high energies. In the optical the delay is 
${\sim}10$
days as noted for the same period of observations \citep{brocksopp06}. The horizontal lines indicate the time intervals 
(one revolution each) over which spectra were obtained. 
The optical/soft X-rays light curve behaved very differently from that of the hard X-rays; this might suggest that the X-ray light curve 
is actually a composite of the two known spectral components, one gradually increasing with the optical/soft X-rays emission 
(accretion disk) and the other following the behaviour of the hard X-rays (jet and/or corona). This is another point 
favouring the unification model explained before (proposed by \citet{esin97}). \citet{brocksopp06} propose that 
during this outburst a transition occurs from LH to HS state.

%PONER FIGURA DEL CAMPO Y EXPLICAR LAS FUENTES QUE ESTAN TAMBIEN ACTIVAS EN EL MISMO PERIODO que las vi en las imagenes
%de ISGRI para analizar los datos de SPI (a pie de figura)!!!!

%PARA RELLENAR ESTA SECCION, EXPLICAR TAMBIEN CADA PUNTO DE LA CURVA DE LUZ QUE REPRESENTA (no para proceeding), I.E. UN PTO.
%UNA SCW O QUE OTRA COSA?

\section{Spectral Analysis}

\begin{table}
  \begin{center}
    \caption{Table with some information of interest of this very preliminar spectra fitting (${\nu}$ means the number of degrees of freedom of the spectral 
fitting).}\vspace{1em}
    \renewcommand{\arraystretch}{1.2}
    \begin{tabular}[h]{llll}
      \hline
      Rev. & Model & ${\chi}^2/{\nu}$ &  Relevant par.\\
      \hline
      0290  & abs${\times}$plw &  114/48 & ${\Gamma}=1.63$ \\
      0295  & abs(bb+cTT) & 163/140 & $T_{in}=1.2{\pm}0.8$ keV \\
            &             &         &  $T_{e}=279{\pm}77$ keV  \\
      0296  & abs(bb+cTT) & 175/126 & $T_{in}=1.3{\pm}0.5$ keV \\
            &             &         & $T_{e}=67{\pm}45$ keV \\
      0299  & abs(bb+cTT) & 584/84  &                            \\
      0304  & abs(bb+cTT) & 609/83  &    \\
      \hline \\
      \end{tabular}
    \label{spect}
  \end{center}
\end{table}

\begin{figure*}
\centering
\vspace{4cm}
\includegraphics[width=0.2\linewidth,angle=-90]{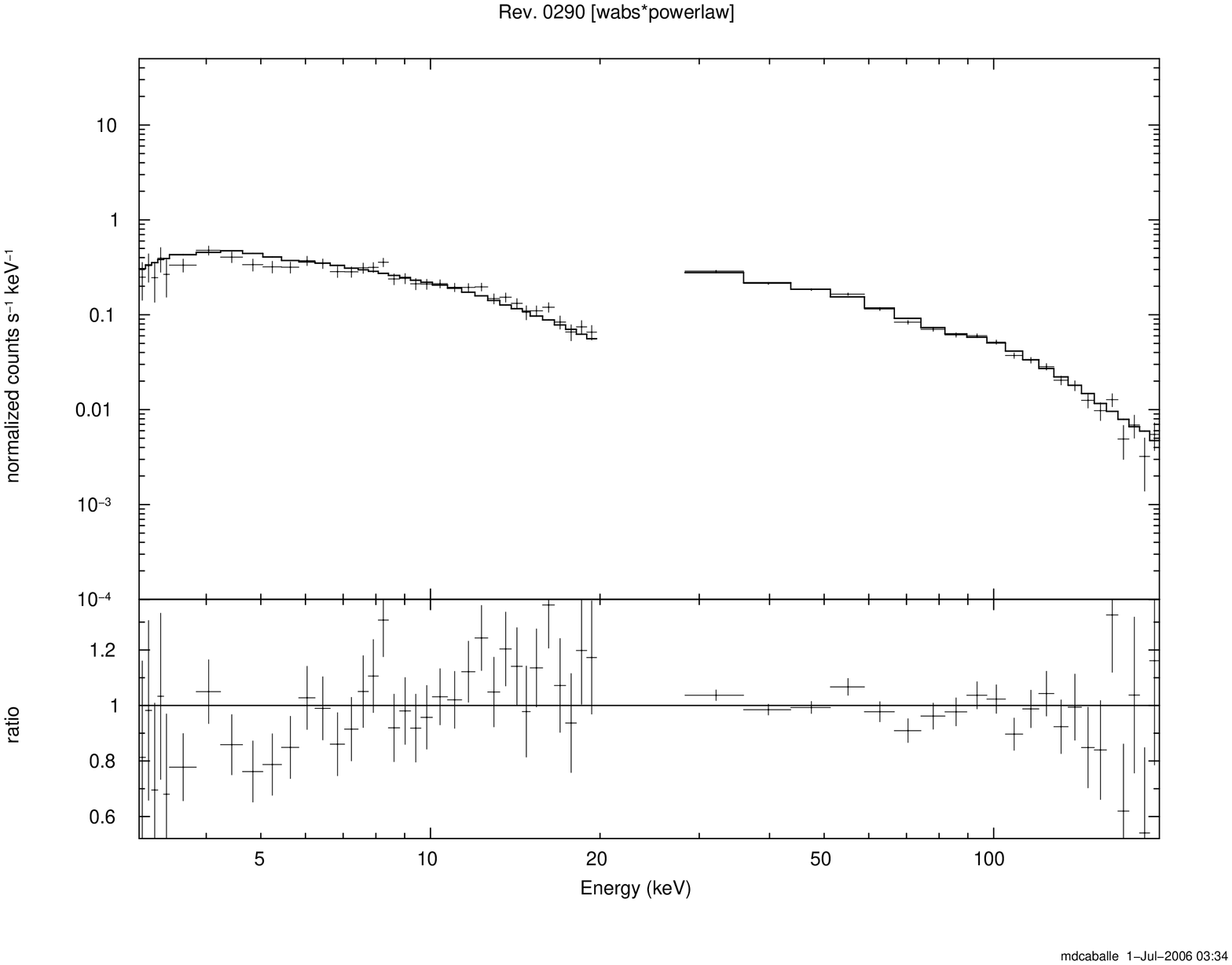}
\includegraphics[width=0.2\linewidth,angle=-90]{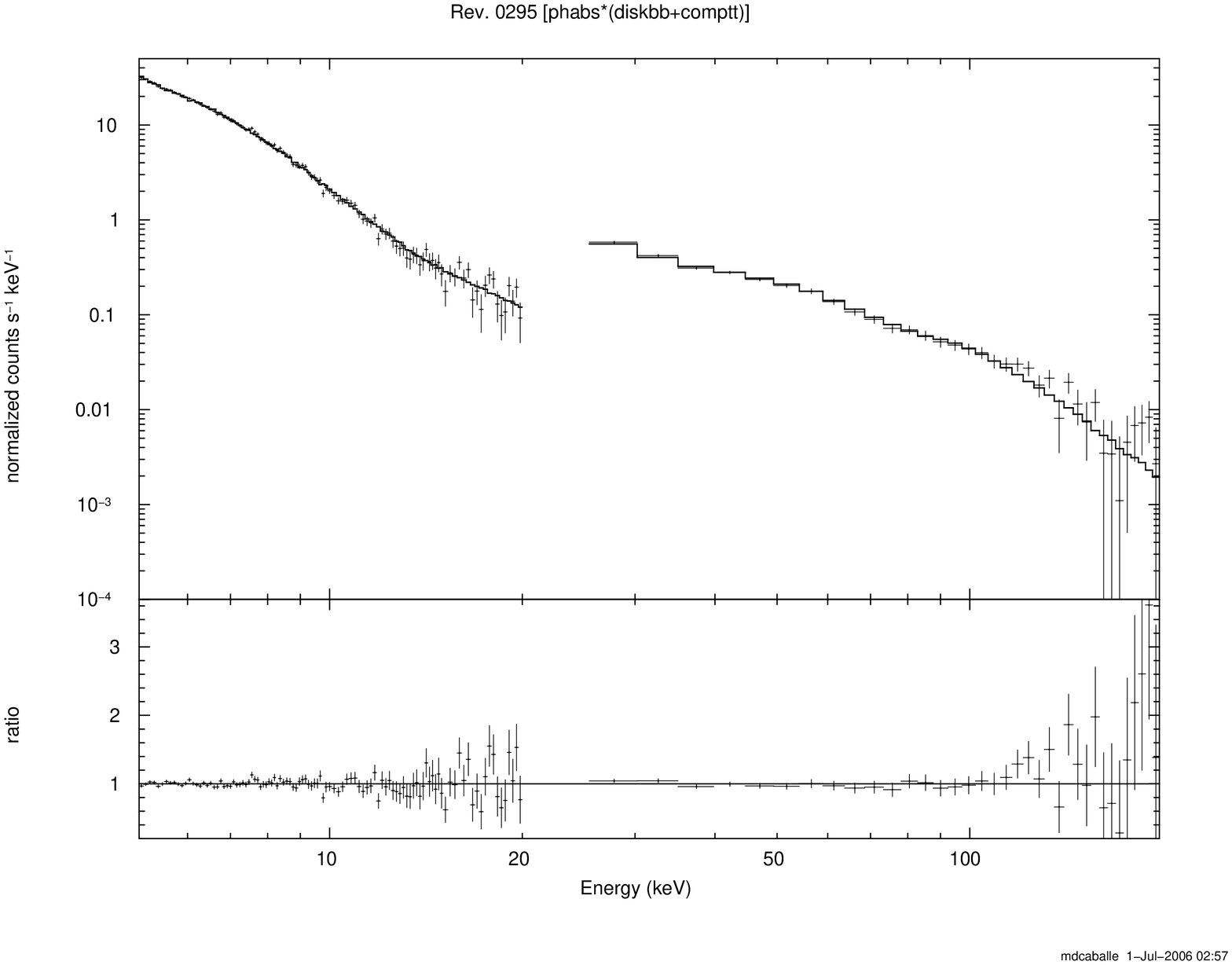}
\includegraphics[width=0.2\linewidth,angle=-90]{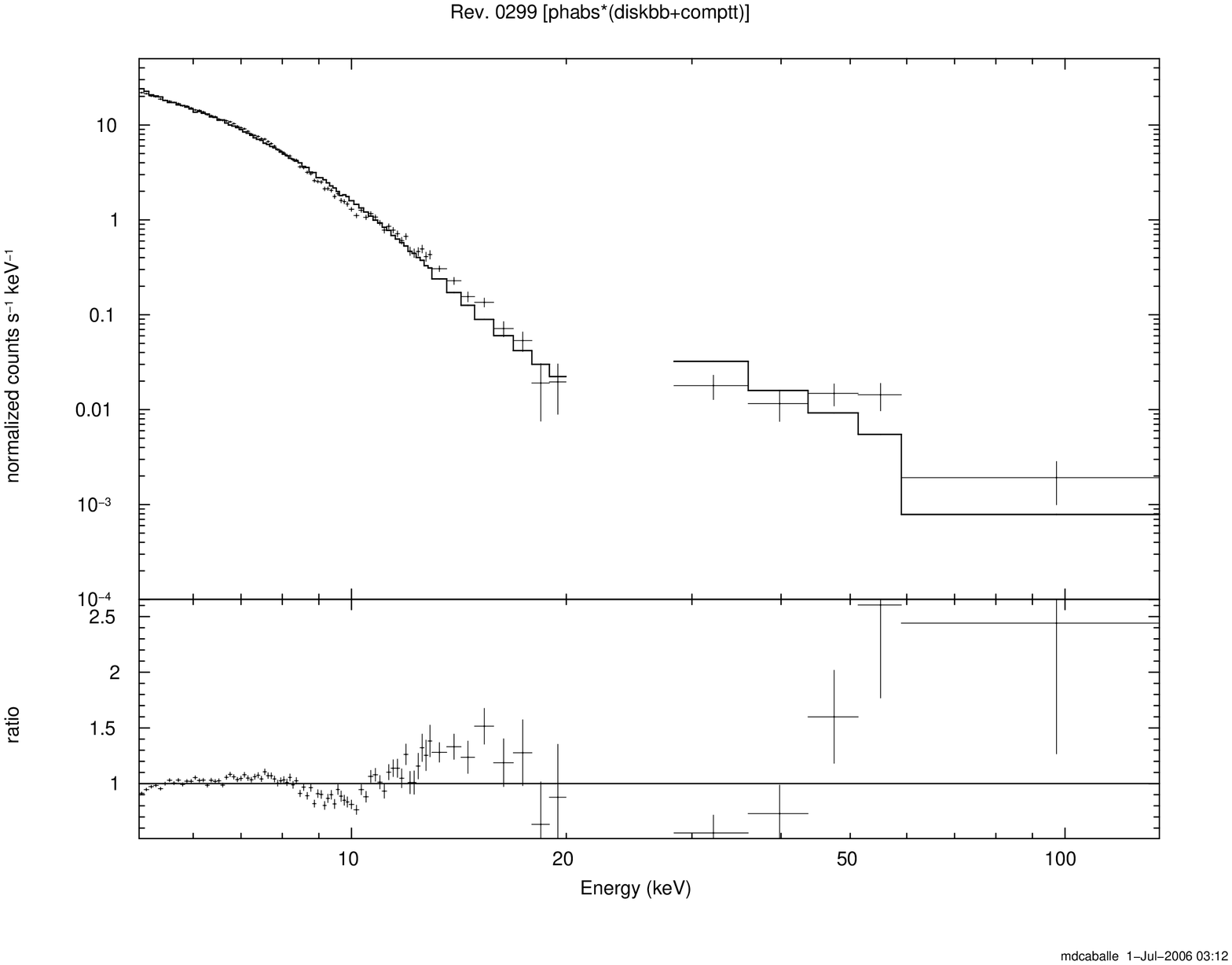}
\caption{Spectra obtained by JEM-X and IBIS/ISGRI on board INTEGRAL for revolutions 290, 295 and 299 (left to right panel). \label{spec}}
\end{figure*}

We performed spectral analysis of the JEM-X and IBIS/ISGRI data.
For JEM-X and ISGRI, individual spectra were obtained for each ScW. The spectra were then combined to obtain an averaged spectrum per revolution (i.e. INTEGRAL revolution 290, 295, 296, 299 and 304) using the $spe\_pick$ OSA tool. This was done, since there was not a significant evolution of the spectra during a revolution and it improved the signal to noise.
%EXPLICAR QUE EN EL CASO DE SPI NO ES POSIBLE HACER ESTO DEBIDO A SU BAJA SENSIBILIDAD (DEBIDO A SU NUMERO REDUCIDO DE 
%DETECTORES), ESTO PARA EL PAPER. 
We then performed a simultaneous fit to the JEM-X and IBIS/ISGRI data, for each of the five revolutions using XSPEC v.12.

In Figure \ref{spec} we show spectra from revolutions 290, 295 and 299 because these are the most significant to see the
evolution of the source during the total period of observations. In Table \ref{spect} we summarize some information obtained
during these very preliminary fits.
%PONER EN EL PAPER LA VERSION UTILIZADA DE XSPEC (11.3.2 en lugar de la 12). 

From this analysis, we observe a strong evolution of the spectrum from a pure hard power law, in revolution 290,
to a model constituted by a photo-absorbed accretion disk bathed in a comptonizing corona (\citet{titarchuk94}), in revolutions
295 and 296. This is accompanied by a strong increase of the photon index (from ${\Gamma}{\sim}1.73{\pm}0.02$, in revolution 290, to
${\Gamma}{\sim}2.2{\pm}0.4$, in revolution 299, the source is almost not detected by ISGRI in revolution 304) together with an increase of the input photon temperature (from $T_{in}{\ll}1$ keV,
in revolution 290, to $T_{in}=1.2{\pm}0.8$ keV, in revolution 295, and $T_{in}=1.3{\pm}0.5$ keV, in revolution 296). 
All together would indicate an increasing of the input photon flux coming from the 
innermost regions of the accretion disk, which are scattered with the electrons present in the hot corona \citep{titarchuk94}.
These features are as expected in the case of successive approaching of the inner disk radius down to the ISCO, as predicted
by the unified model from \citet{esin97}. In revolutions 299 and 304 the photo-absorbed and comptonized disk model is also better than photo-absorbed disk alone, but a more complex spectrum (which requires reflection and likely a weak iron line) appears to be more appropiate. 

%PONER AQUI LOS MODELOS AJUSTADOS Y UN POCO DE LOS PARAMETROS OBTENIDOS (no mucho pq es el proceeding)

\section{Discussion}

As we noted above, there was a transition from an LH state to an HS state \citep{brocksopp06}. 
This can be explained in the common accepted view of black hole X-ray transients, in 
which the low-mass companion star undergoes Roche-lobe overflow, but at a sufficiently low rate that the gas is accumulated
in the (cool) disk until a critical level is reached, at which point the outburst occurs (\citet{lasota01}; \citet{meyer04}).
In between outbursts, the accretion rate is low and is conjectured that
an ADAF model region fills the inner accretion disk. 
%PONER LO QUE A MI ME HA SALIDO!!!! 
With this scenario, in the quiescent and low 
states the accretion is low and the spectrum is hard as is shown in our spectrum from revolution 290.
During the outburst, the spectrum changes to a ${\sim}1$keV thermal spectrum, often interpreted as
the radiation from a geometrically thin, optically thick accretion disk, superposed on a softer power law (photon indices 
${\Gamma}{\sim}2-3$) extending to ${\sim}200$ keV, as shown in our spectra from revolutions 295 and 296.
When the accretion disk becomes thick and very near to the ISCO, then the reflection features are manifested in the shape
of the spectra (see \citet{done01} and references therein), as can be seen in our spectra from revolutions 299 and 304.
It is worth noticing that, when matter is close to the BH, jet production is an alternative mechanism for the origin of the hard X-rays.

We can conclude that our spectra is consistent with major models for outburst evolution, including model of \citet{esin97}.

%ESTUDIAR MAS LOS MODELOS DE COMPTONIZACION, Y CUANDO SUELEN APARECER (ESTO PARA EL PAPER).

\section*{Acknowledgments}

This research is partially supported by Spanish MEC under
grants PNE2003-04352+ESP2005-07714-C03-03.

% The following bibliography was produced with
%   \bibliographystyle{aa}
%   \bibliography{esapub}
% The results are inserted directly here to simplify
% the demonstration.

\end{document}